\documentclass[preprint,preprintnumbers,amsmath,amssymb]{revtex4}
\usepackage{graphicx,epstopdf}

\usepackage{graphicx}% Include figure files
\usepackage{dcolumn}% Align table columns on decimal point
\usepackage{bm}% bold math
\usepackage{amssymb}

\begin{document}
\title{Control of optical emission in doped GaAs/AlGaAs nanofabricated quantum dots}
\author{Sokratis Kalliakos, C\'{e}sar Pascual Garc\'{i}a, Vittorio Pellegrini}
\affiliation{NEST CNR-INFM and Scuola Normale Superiore, Piazza dei
Cavalieri 7, I-56126 Pisa, Italy}
\author{Marian Zamfirescu, Lucia Cavigli, Massimo Gurioli, and Anna Vinattieri}
\affiliation{Department of Physics and LENS, Universit\`{a} di Firenze, 50019 Sesto Fiorentino, Italy}
\author{Aron Pinczuk*, Brian S. Dennis, Loren N. Pfeiffer, Ken W. West}
\affiliation{Bell Labs, Alcatel-Lucent, Murray Hill, New Jersey}
\affiliation{*Dept of Physics, Dept of Appl.~Phys.~and
Appl.~Math., Columbia University, New York, New York}
\date{\today}
\begin{abstract}
Dilute arrays of GaAs/AlGaAs modulation-doped quantum dots (QDs)
fabricated by electron-beam lithography and low impact
reactive-ion etching exhibit highly homogeneous luminescence.
Single quantum dots display spectral emission with peak energies
and linewidths linked largely to the geometrical diameter of the
dot and to the built-in electron population.  Excitonic-like and
biexcitonic-like emission intensities have activation energy of
about 2 meV. These results highlight the potential of high quality
nanofabricated QDs for applications in areas that require fine
control of optical emission.
\end{abstract}
\maketitle
\par
Semiconductor QDs are the materials basis for the realization of
high-performance optoelectronic devices \cite{review, finley}. QDs
have been also proposed as building blocks for spintronic devices
and for the creation and manipulation of quantum bits at the
nanoscale \cite{loss}. In addition, they represent a unique
laboratory where few-particle fundamental quantum effects can be
studied \cite{garcia}. A large part of the spectroscopic
investigations of QDs was devoted to undoped self-assembled QDs
that however offer limited control on their size and emission
energies. Their macro-photoluminescence spectra exhibit large
inhomogeneous broadening due to the size variation from dot to dot
\cite {gurioli}. In single-dot micro-photoluminescence ($\mu$-PL)
experiments, however, sharp homogeneous-broadened excitonic
emission peaks were reported \cite{finley2,dekel}. Single-dot
luminescence was also carried out in undoped AlGaAs/GaAs QDs
formed by well-width fluctuations or defined in cleaved-edge
overgrowth samples \cite{weg}. Nanofabricated modulation-doped
GaAs/AlGaAs QDs populated by many electrons were also proposed and
studied by inelastic light scattering \cite{strenz}. Recently we
demonstrated that arrays of nanofabricated GaAs/AlGaAs QDs doped
with very few electrons can be fabricated and studied by inelastic
light scattering spectroscopy \cite{garcia}.
\par
In this letter we address the interband optical properties of
few-electron nanofabricated GaAs/AlGaAs QDs by micro-luminescence.
We found that their emission energies can be tuned by changing the
QD metallurgical diameter and that their emission lineshapes are
identical in different individual QDs. By performing a
power-dependent analysis of the spectra we have determined
excitonic-like and biexcitonic-like recombinations, even in the
presence of a built-in electron population. These excitonic peaks
show remarkable wide line-widths close to 1 meV contrary to what
is usually observed in single undoped QDs. Finally we derived an
activation energy of about 2 meV from the analysis of the
temperature dependence of the intensities of excitonic
luminescence. The results highlight the robust and reproducible
tunable optical recombination properties of the high quality
nanofabricated structures. Access to such uniform QD systems also
offer experimental platforms for the explorations of fundamental
interactions in inter-band optical properties of QDs charged with
electrons \cite{warburton, wojs},
\par
Arrays of QDs were fabricated by electron beam lithography and
reactive ion etching using a single modulation-doped quantum well
(QW) of 25nm \cite{garcia}. Scanning Electron Microscope (SEM)
pictures of one array as well as of individual dots are shown in
Fig. 1 (b,c,d). Here we focus on QDs with diameter $d$ in the
range 210-440 nm for which the electron occupation ranges from
around  4-6 to few tens. We remark that in the case of GaAs QDs
defined by dry etching, the number of confined electrons and the
effective confinement are both determined by the large depletion
region associated to Fermi level pinning due to GaAs surface
states such as antisite defects \cite{spicer} or other chemical
defects introduced during the etching process. Due to the
depletion effect, the effective sizes of the electronic
confinement are much smaller than the geometrical diameters $d$ of
the etched mesa structures particularly at low-electron densities
\cite{martorell, garcia}.
\par
For the $\mu$-PL experiments, a dye laser at 600 nm pumped by the second harmonic of a Nd:YAG pulsed laser was used with pulse duration of 2 ps and repetition rate of 76 MHz. For the collection, a confocal configuration of two infinity corrected microscope objectives was used. The PL signal was then focused to a monomode optical fiber with a core diameter of 3.5 $\mu$m assuring a lateral resolution of 0.7 $\mu$m. A single grating spectrometer and a CCD camera were used for the detection. All $\mu$-PL experiments were carried at 5.7K (except for the temperature-dependent spectra) in a low-vibrational cryostat. The presence of an electron population in the dots was confirmed separately by inelastic light scattering experiments displaying a rich spectrum of intershell electronic excitations in all arrays.
\par
Figure 1(a) shows the $\mu$-PL spectra of a single QD with $d$=280 nm. The homogeneity of the optical emission of the QDs was inferred by performing spatially-resolved PL experiments along the arrays that revealed identical spectra from different excited QDs leading to a macro-PL very similar to the $\mu$-PL spectrum. The peak at $\approx $ 1.525 eV is assigned to the QD PL and it is blue-shifted with respect to the QW peak as shown in Fig.2 where the $\mu $-PL emission peak of the $d$=330 nm QD is also reported. The QW emission presents the usual lineshape of modulation-doped QW luminescence determined by the recombination of electrons from the bottom of the subband up to the Fermi energy ($E_F \approx $ 2 meV). The single QD emissions are well described by Lorentzian lineshapes (shown as red lines in Figs.1(a), 2 and 3(a)) suggesting homogeneous broadening as expected for single-dot luminescence. In addition, the plot of the integrated intensity of the peak shown in Fig.1(a) with temperature changes yields an activation energy of $E_a = 2\pm 0.65 $ meV, which is deduced from the simple equation $I$ = $A$ - $B$exp$(-E_a /K_B T)$ (solid line in inset to Fig.1(a))\cite{pinczuk}. This value, which is comparable to the QD inter-shell energy difference \cite{garcia}, suggests that thermal activation of excited excitonic states can be responsible for the quenching of the PL signal. The inset to Fig. 2 summarizes the energy shift of the QD PL with respect to the QW PL peak (at 1.521 eV) for different dot diameters. The observed red-shift with increasing $d$ reflects the decrease of the effective three-dimensional confinement with confinement energies of the order of 1 meV already observed for $d$ = 440 nm due to the impact of the depletion region. Full widths at half maximum (FWHM) of the emission peaks are much larger than those usually observed in single-dot PL with values of 0.7 meV for the $d$ = 280 nm dot and 2.7 meV for the $d$ = 330 nm QD ($\approx $2 meV for the $d$ = 440 nm QD). Finally we note that the small higher energy shoulder shown in Fig.1(a) is observed at the same energy position in other QD arrays with different diameters and it is assigned to recombination from the Si-doped AlGaAs layer, probably linked to deep donor-acceptor pairs \cite{nota}.
\begin{figure}
\includegraphics[width=9cm]{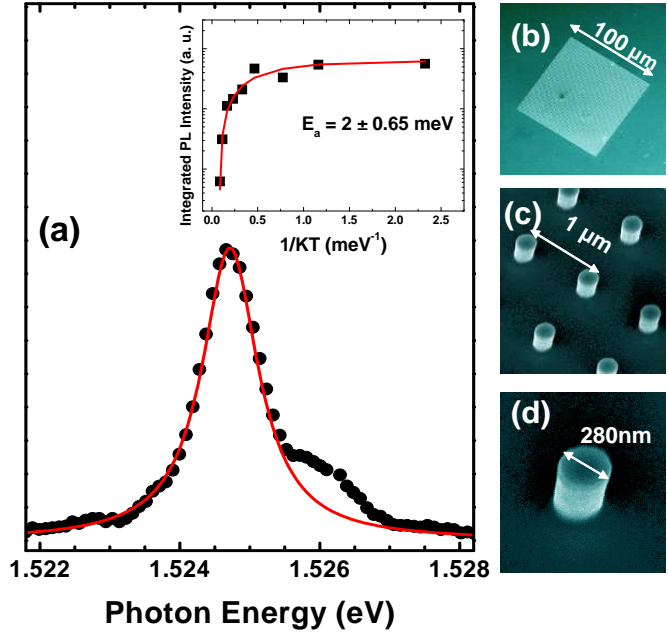}
\caption{\label{fig1} (color online) (a)$\mu$-PL spectrum (filled circles) of a single dry-etched quantum dot (QD) of geometrical diameter $d$=280 nm at T=5.7K. The power density is $5 x 10^{-3} W/cm^2$. The red line shows the result of a fitting analysis with a Lorentzian function. Inset: Plot of the integrated PL intensity with temperature changes (black squares). The red line corresponds to an activation energy of 2 meV. (b) SEM image of the $100\mu m$ x $100\mu m$ array composed by $10^{4}$  QDs replica (c) of QDs separated by $1 \mu m$ and (d) of one dry-etched QD.}
\end{figure}
\par
In order to identify the nature of the PL emission shown in Fig. 1(a) we report in Fig. 3(a) the power-dependence of the $\mu $-PL. As the power density increases, an additional peak appears as a shoulder in the low-energy side of the main peak, and red-shifted by 1.5 meV. A fitting procedure using two Lorentzian lines reproduce very well the observed peaks as demonstrated in Fig. 3(a). The inset to Fig. 3(a) shows the integrated intensities of the main and the low-energy QD peaks as a function of power density. The main QD peak (peak A) has a linear dependence with power density, a typical excitonic-like behavior. On the contrary, the lower-energy peak (peak B) has a quadratic dependence with the power density, which is a typical signature of biexcitonic-like behavior \cite{senellart}. The 1.5 meV energy shift between these two peaks can thus be interpreted as the binding energy of the biexcitonic excitation in the presence of the electron population. The temperature-dependent evolution of the spectra shown in Fig. 3(b) demonstrates that beyond 20K the biexciton disappears which is consistent with a binding energy of 1.5 meV.
\begin{figure}
\includegraphics[width=9cm]{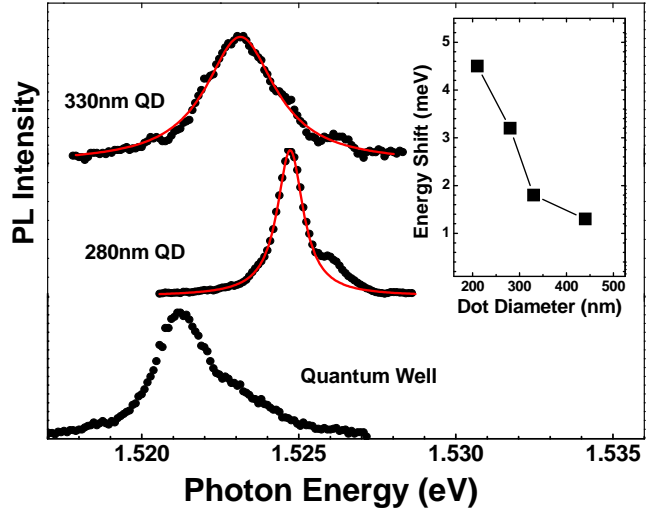}
\caption{\label{fig2} (Color online) $\mu$-PL spectrum of a single quantum dot of geometrical diameter $d$=330 nm (upper spectrum), $d$= 280 nm (middle spectrum) and of the quantum-well emission (lower spectrum) at T=5.7K.  Red lines show the results of a fitting analysis with a single Lorentzian function. Inset: Energy shift of the QD emissions from the quantum-well emission for different QD diameters $d$.}
\end{figure}
\par
The evolution of the energy shifts of the QD peaks from the QW emission ranges from 1 meV for the $d$=440 nm dot up to 4.6 meV for the $d$=210 nm dot (see inset to Fig.2) and allows to give an estimate for the number of electrons in the dots.
The QD emission energy can be written as \cite{wojs} $E^{QD} = E^{QW} + \hbar \omega_e + \hbar \omega_h +E^{QD}_{ee}$, where $\hbar \omega_e$ and $\hbar \omega_h$ are the in-plane confinement energies for the electrons and holes, respectively, $E^{QW}$ is the QW transition energy and $E^{QD}_{ee}$ measures many-body corrections due to the QD confinement. $E^{QD}_{ee}$ is negative and determined by the excitonic attraction of the QD electron-hole pair and the interaction between the pair with the excess electrons in the QD.
\par
For QDs with $d$=210 nm, $\hbar \omega_e = 4 meV$ and $N\approx 4$ ($N$ is the electron population in the QD) were previously determined \cite{garcia}. In the parabolic approximation we thus have $\hbar \omega_h = (m_e/m_h)^{0.5} \cdot \hbar \omega_e = 3.1 meV$ where $m_h = 0.11m^*$ is the in-plane heavy-hole effective mass in the cylindrical approximation. The measured energy shift $E^{QD}-E^{QW}$= 4.5 meV thus implies that $E^{QD}_{ee}\approx 2.6 meV$. If we assume the same value of $E^{QD}_{ee}$ for the QDs with $d$=280 nm we obtain $\hbar \omega_e \approx 3.25 meV$ that leads to $N\approx 10$ \cite{reim}. In the case of QDs with larger diameter, we expect a significant reduction of $E^{QD}_{ee}$ together with the deviation of the potential profile from the parabolic form, due to the increased number of electrons. It is therefore difficult to estimate precisely the number of electrons but we expect an electron occupation that approaches a few tens for the largest QDs with $d$ = 440 nm.
\par
\begin{figure}
\includegraphics[width=8cm]{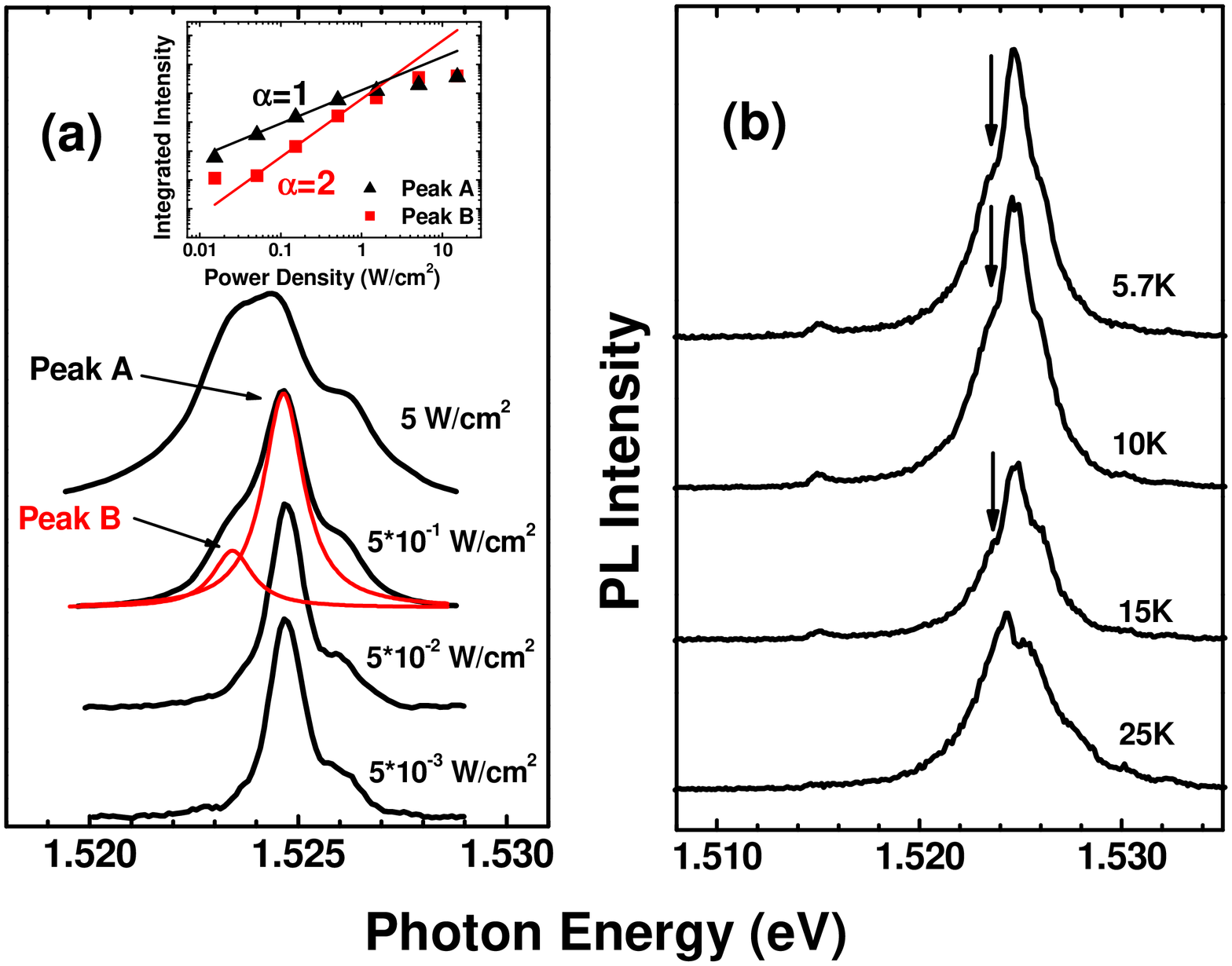}
\caption{\label{fig3}  (Color online)(a):  $\mu$-PL spectra of a single dry-etched quantum dot of geometrical diameter $d$ = 280 nm as a function of power density. The red lines are the two Lorentzian lines used for the fitting analysis. Inset: integrated intensity of the high energy ($A$ - black triangles) and low energy ($B$ - red squares) peaks as a function of the power density. The solid lines represent the linear ($\alpha=1$, black line) and quadratic ($\alpha=2$, red line) intensity dependence on the power density. (b) Temperature dependence of the $\mu$-photoluminescence spectrum of a single 280nm quantum dot. The arrow indicates the energy position of the biexcitonic peak. The power density is $5x10^{-1} W/cm^2$.}
\end{figure}
\par
As mentioned before, the linewidth of the (Lorentzian) peaks in our spectra is remarkably large for single-dot spectroscopy. Spectral diffusion does not seem to be the responsible broadening mechanism since one would expect its impact to be reduced in larger dots (which is not the case). Instead the broadening could reflect the dynamical changes of the electron population induced by the interband electron-hole recombination. A similar process was considered in Ref. \cite{warburton} to explain the increase of the $\mu$-PL linewidth in doped InAs quantum rings at electron occupation beyond $N$=4. Further investigations are needed to establish the correct broadening mechanism in these systems.
\par
In conclusion, we studied the $\mu$-photoluminescence of
nanofabricated charged AlGaAs/GaAs single QDs. Excitonic and
biexcitonic-like recombinations are observed, even in the presence
of a built-in electron population with recombination energies
linked to the QD lateral diameters. The nanofabricated GaAs/AlGaAs
QDs with a limited and controlled number of electrons show
promises for applications in quantum optics as well as for
fundamental studies of electron correlations in nanoscale systems.
\par
We are grateful to M. Rontani, G. Goldoni and E. Molinari for
illuminating discussions. We acknowledge support from the Italian
Ministry of Foreign Affairs, Italian Ministry of Research
(FIRB-RBAU01ZEML), European Community's Human Potential Program
(HPRN-CT-2002-00291). AP is supported by the National Science
Foundation (DMR-03-52738), the Department of Energy
(DE-AIO2-04ER46133), the Nanoscale Science and Engineering
Initiative of the National Science Foundation under NSF Award
Number CHE-0117752, and the New York State Office of Science,
Technology, and Academic Research (NYSTAR).

\end{document}